\documentstyle[pra,aps]{revtex}

\def\be{\begin{equation}}
\def\ee{\end{equation}}

\def\binom#1#2{{\left({#1 \atop #2}\right)}}

\begin{document}
\draft
\title{Concurrence in arbitrary dimensions.}
\author{Piotr Badzi{\c{a}}g$^{1,}$\cite{poczta}, 
Piotr Deuar$^{3,}$\cite{pocztaD}, \\
Micha\l{} Horodecki$^{2,}$ \cite{poczta1}, Pawe\l{} Horodecki$^{3,}$\cite
{poczta2}, and Ryszard Horodecki $^{2,}$\cite{poczta3} }

\address{
$^{1}$ Department of Mathematics and Physics , \\
M\"{a}lardalens H\"{o}gskola, S-721 23 V\"{a}ster\aa s, Sweden, \\
$^{2}$ Institute of Theoretical Physics and Astrophysics,\\
University of Gda\'nsk, 80--952 Gda\'nsk, Poland,\\
$^{3}$Faculty of Applied Physics and Mathematics,\\
Technical University of Gda\'nsk, 80--952 Gda\'nsk, Poland\\
}
\maketitle

\begin{abstract}
We argue that a complete characterisation of quantum
correlations in bipartite systems of many dimensions may require a quantity
which, even for pure states, does not reduce to a single number.
Subsequently, we introduce multi-dimensional generalizations of concurrence
and find evidence that they may provide useful tools for the analysis of
quantum correlations in mixed bipartite states. We also introudce 
{\it biconcurrence} that leads to a necessary and sufficient condition 
for separability.
\end{abstract}

\section{Introduction.}

Entanglement plays central role in quantum information theory \cite{xxx1}.
Pure state entanglement of bipartite systems is well understood in the sense
that the relevant parameters for its optimal manipulation by local
operations and classical communication (LOCC) have been identified and
analyzed \cite{Bennet1}, \cite{Nielsen+}. Many efforts have also been
devoted to the study of mixed-state entanglement. There, several possible
entanglement measures have been proposed. Among these, entanglement of
formation ($E_{F}$) \cite{Bennet-mixed}, \cite{HaydenMHBT} attracts much of
attention, as it is closely connected with (or, perhaps, equal to) the rate
of production of mixed bipartite states out of pure ones by LOCC operations.
It is, however, extremely difficult to evaluate $E_F$, but for the
analytical formula for a single copy of an arbitrary state of two qubits
obtained by Wootters \cite{Wootters}. Despite efforts, not much progress has
been recorded regarding generalization of Wootters' result to the states in
more than $2\times2$ dimensions \cite{Werner}.

Wootters' success in quantifying $E_{F}$ for two qubits can be attributed to
associating $E_{F}$ with concurrence which is easier to calculate than $%
E_{F} $. Concurrence, as introduced by Hill and Wootters \cite{Hill&Wootters}%
, was defined via operation of spin flip. More recently, Rungta et al. \cite
{Rungta}\ made an attempt to generalize the notion of concurrence to pure
bipartite states in arbitrary dimensions by introducing operation of
universal state inversion \cite{zero}. Their universal inverter generalizes
spin flip to a transformation which brings pure state $|\psi \rangle $ into
the maximally mixed state in the subspace orthogonal to $|\psi \rangle $. In
the same way that the spin flip generates concurrence for a pair of qubits,
the universal inverter generates a number which generalizes concurrence for
joint pure states of pairs of quantum systems of arbitrary dimensions.
Generalized in this way, concurrence measures entanglement of pure bipartite
states in terms of the purity of their marginal density operators.

As one knows \cite{Nielsen+}, a complete characterization of quantum
correlations in bipartite systems of many dimensions may require a quantity
which, even for pure states, does not reduce to a single number \cite{asym}.
Take, e.g., two pure states represented by vectors$\ \psi =\left( |11\rangle
+|22\rangle \right) /2$ and $\phi =a\psi +b|33\rangle $, with $a=\sqrt{x}$\
and $b=\sqrt{1-x}$, where $x\approx 0.2271$ is a root of $x^{x}\left[
2\left( 1-x\right) \right] ^{1-x}=1$. The two states have the same
entanglement $E_{F}$ of 1 ebit, nevertheless they have different Schmidt
numbers and, consequently, it is impossible to locally convert one into the
other.

In this contribution, we argue than that a suitable generalization of spin
flip to more dimensions should produce a multi-dimensional analogue of
concurrence rather than a single number. Such a concurrence would then
describe not only the amount of entanglement but also its structure, e.g.,
the size (the number of dimensions) of the entangled spaces on each side.
The concurrence for pure states is thus a matrix acting on the antisymmetric
subspace of the total Hilbert space of two systems. Having that, one can
follow Wootters and generalize the concept to mixed states by introducing a
matrix of preconcurrence. The elements of this matrix are matrices in their
own right and at the end, the matrix is often difficult to analyze. At least
partially, the difficulties can be associated with the matrix dependence on
the choice of the local bases. Therefore, we also generalize the concept of
concurrence in a somewhat different direction. We abandon the requirement
for preconcurrence to be a second order object in the state's ensemble. For
this price we can define a fourth order object, \textit{biconcurrence}
matrix. It is independent of the local unitaries and allows us to
reformulate separability problem in terms of the main diagonal of the
matrix. Biconcurrence is a very simple function of ensemble of density
matrix and has many symmetries. Therefore, the obtained necessary and
sufficient separability condition seems to be the most promising one from
algebraic point of view.

The generalization of pre-concurrence which satisfies our criterion is
presented in section \ref{sec-spinflip}. Then, in section \ref{sec-concsep}
we give an example to show how our multi-dimensional pre-concurrence can be
used for analysis of separability in arbitrary dimensions. There, we also
discuss possible limitation of such analysis. Subsequently, in Sec. \ref
{sec-biconcurrence} we introduce biconcurrence and formulate the necessary
and sufficient condition for separability in terms of its elements. Finally,
in section \ref{sec-conclusion} we present a brief discussion of the results.

\section{Spin flip and concurrence.}

\label{sec-spinflip}

\subsection{Pure states.}

For pure states of two qubits, a spin flip transforms vector $v$ in a
2-dimensional vector space into vector $\tilde{v}$\ equally long and
orthogonal to $v$. In a bipartite system, a spin flip means that Alice
performs a spin flip on her qubit and Bob on his. This gives a particularly
simple expression for concurrence: 
\begin{equation}
C\left( \psi \right) =\langle \tilde{\psi}|\psi \rangle   \label{Con-pure1}
\end{equation}
The spin flip operation and the concurrence which follows are well defined
since, in a 2-dim space, there is only one direction which is orthogonal to
a given one. One may further notice that concurrence defined in (\ref
{Con-pure1}) together with the state's normalization allow to determine the
eigenvalues of the associated reduced density matrix and, via these, the
pure state's entanglement. The eigenvalues are the squares of the moduli of
the singular values $\lambda _{1}$ and $\,\lambda _{2}$ of a $2\times 2$
matrix $\left[ \psi \right] $\ of the coefficients defining the state in the
standard basis: 
\begin{equation}
\left| \psi \right\rangle =\sum\limits_{i,j}\psi _{i,j}\,\left|
i\right\rangle _{A}\otimes \left| j\right\rangle _{B}  \label{pure-state}
\end{equation}
\ The singular values are then related to the concurrence via 
\[
C=2\,\lambda _{1}\,\lambda _{2}=2\,\det \left( \left[ \psi \right] \right) 
\]
In general, in a \emph{d}-dimensional space there are $d-1$ dimensions
orthogonal to a given direction. These can be represented by a $d-1$
antisymmetric form. From this point of view, performing a spin flip on a
bipartite state means constructing a double $d-1$ form (one side for Alice
and one side for Bob) locally dual to the double one-form representing the
state vector $\left| \psi \right\rangle $. Concurrence can then be
associated with the contraction of the form representing $\left| \psi
\right\rangle $ with the form representing $\left\langle \tilde{\psi}\right| 
$. The contraction gives a double $\left( d-2\right) $-form which is
equivalent to a double $2$-form and can be represented by a 
$\binom{d}{2}%
\times \binom{d}{2}$\ matrix with the following elements: 
\begin{equation}
C_{i_{1}\wedge j_{1};i_{2}\wedge j_{2}}=2\left( \psi _{i_{1},i_{2}}\,\psi
_{j_{1},j_{2}}-\psi _{i_{1},j_{2}}\,\psi _{j_{1},i_{2}}\right) 
\label{Con-pure2}
\end{equation}
These elements are easily identified as twice the two-dimensional minors of
matrix $[\psi ]$. They describe the two-state contributions to the bipartite
entanglement.

Regarding their structure, matrices $C$ form a vector space with a natural
trace norm: 
\begin{equation}
\left| C\right| ^{2}=Tr\left( C\,C^{\dagger }\right) =\sum\limits_{i\wedge
j,k\wedge l}\left| C_{i\wedge j;k\wedge l}\right| ^{2}  \label{Con-pure norm}
\end{equation}
Having constructed the concurrence matrix, one may proceed in the same
spirit and construct higher dimensional minors of $[\psi ]$ (up to the
Schmidt number). They will represent those contributions to the bipartite
entanglement which embrace local subspaces of higher dimensions. We believe
that, in principle, these concurrences of order higher than two may be
important for the quantification of entanglement even if the separability of
a pure state is determined by the lowest order (i.e. 2) concurrence.
Clearly, a pure state (\ref{pure-state}) in arbitrary dimensions is
separable iff $\left[ C\right] =0$.

\subsection{Mixed states.}

\label{subsec-mixed-states} In order to further generalize the concept of
concurrence to multidimensional mixed states, we follow Wootters and
introduce (pre)concurrence as follows. Given a decomposition of state $%
\varrho $ into pure, unnormalized states, 
\begin{equation}
\varrho =\sum\limits_{\mu }\left| \psi ^{\mu }\right\rangle \left\langle
\psi ^{\mu }\right|   \label{state}
\end{equation}
we define pre-concurrences 
\begin{equation}
\begin{array}{ccc}
C_{i_{1}\wedge j_{1};i_{2}\wedge j_{2}}^{\mu \nu } & = & \psi
_{i_{1},i_{2}}^{\mu }\,\psi _{j_{1},j_{2}}^{\nu }-\psi _{i_{1},j_{2}}^{\mu
}\,\psi _{j_{1},i_{2}}^{\nu } \\ 
&  &  \\ 
& + & \psi _{i_{1},i_{2}}^{\nu }\,\psi _{j_{1},j_{2}}^{\mu }-\psi
_{i_{1},j_{2}}^{\nu }\,\psi _{j_{1},i_{2}}^{\mu }
\end{array}
\label{Con-mixed}
\end{equation}
The pre-concurrences can be regarded as a set of $\binom{d}{2}\times \binom{d%
}{2}$ matrices in $\mu $ and $\nu $ or, equivalently, as one matrix in $\mu $
and $\nu $ with vector-like elements living in a $\binom{d}{2}\times \binom{d%
}{2}$ dimensional space.

To systematize this picture, it may also be convenient to view $C$ as an
operator in the tensor product of two spaces. The first, $\mathcal{H}_{1}$
is the antisymmetric subspace of the space $C^{d}\otimes C^{d}$ the state
acts on. Thus $\mathcal{H}_{1}=C^{d^{2}-d/2}$. The space $\mathcal{H}_{2}$
is the space of ''lists'' of vectors for decomposition of the state. In
principle we should allow this space to be infinite-dimensional, as one can
consider infinite decompositions. However, it is likely that dimension $%
d^{4} $ is enough (for example, a separable state can be certainly
decomposed into no more than $d^{4}$ product states \cite{Pawel}; similarly, there always
exists an optimal decomposition for entanglement of formation containing no
more than $d^{4}$ components \cite{Uhlmann}).

Matrix $C$ viewed as operator acting on ${\mathcal{H}}_{1}\otimes {\mathcal{H}}%
_{2}$ has simple transformation rules under $(i)$ change of decomposition 
$(ii)$
local unitary transformations of the state. Operations of type $(i)$ 
transform
the preconcurrence matrix according to: 
\begin{equation}
C^{\mu ^{\prime }\nu ^{\prime }}=\sum\limits_{\mu ~\nu }U^{\mu ^{\prime }\mu
}\,C^{\mu \nu }\,U^{\nu ^{\prime }\nu }  \label{Con-sim}
\end{equation}
with $U$ being a unitary matrix changing the decomposition of the state into
pure states \cite{ensembles}. This transformation can be represented as: 
\begin{equation}
C\quad \rightarrow \quad C^{\prime }=I\otimes UCI\otimes U^{T}
\end{equation}
where subscript $T$ stands for transposition. Similarly, a unitary
transformation of the local bases 
\begin{equation}
\left| e_{i_{1}}\otimes f_{i_{2}}\right\rangle
=\sum\limits_{k_{1}~k_{2}}\left| \hat{e}_{k_{1}}\otimes \hat{f}%
_{k_{2}}\right\rangle V_{k_{1}i_{1}}~W_{k_{2}i_{2}}  \label{local-coord}
\end{equation}
(matrices $V$ and $W$ unitary) changes the components of the elements of $%
C^{\mu \nu }$ according to 
\begin{equation}
\begin{array}{ccl}
\mathbf{\hat{C}}_{i_{1}\wedge j_{1};i_{2}\wedge j_{2}}^{\mu \nu } & = & 
\sum%
\limits_{k_{1}l_{1}k_{2}l_{2}}V_{i_{1}k_{1}}V_{j_{1}l_{1}}W_{i_{2}k_{2}}W_{j_{2}l_{2}}~%
\mathbf{C}_{k_{1}\wedge l_{1};k_{2}\wedge l_{2}}^{\mu \nu } \\ 
&  &  \\ 
& = & \sum\limits_{k_{1}<l_{1};k_{2}<l_{2}}\left(
V_{i_{1}k_{1}}V_{j_{1}l_{1}}-V_{i_{1}l_{1}}V_{j_{1}k_{1}}\right) \left(
W_{i_{2}k_{2}}W_{j_{2}l_{2}}-W_{i_{2}l_{2}}W_{j_{2}k_{2}}\right) \mathbf{C}%
_{k_{1}\wedge l_{1};k_{2}\wedge l_{2}}^{\mu \nu }
\end{array}
\label{local-C}
\end{equation}
which can be represented as 
\begin{equation}
C\quad \rightarrow \quad \hat{C}=\left( V\otimes W\right) \otimes
I\,C\,(V^{T}\otimes W^{T})\otimes I
\end{equation}

\section{Concurrence and separability.}

\label{sec-concsep}

The preconcurrence matrix defined in the previous section sheds some
interesting light on the separability of mixed states. Obviously, a given
bipartite state $\rho $ is separable iff there is a decomposition for which
all the diagonal elements $C^{\mu \mu }$\ are zero vectors. The
non-separable states can then be divided into two classes:

\begin{enumerate}
\item[a)]  the states which allow for such a pair of local bases that for at
least one $\kappa _{0}=i_{1}^{0}\wedge j_{1}^{0};i_{2}^{0}\wedge j_{2}^{0}$,
no transformation (\ref{Con-sim}) can zero the diagonal of $C_{\kappa
_{0}}^{\mu \nu }$.

\item[b)]  the states where for every single component $\kappa =i_{1}\wedge
j_{1};i_{2}\wedge j_{2}$, there exist a decomposition with all the diagonal
elements $C_{\kappa }^{\mu \mu }$ to zero (different decompositions for
different multi-indexes $\kappa $). This property must hold irrespective of
the choice of the local bases.
\end{enumerate}

The states in class (a) contain 2-qubit entanglement and as such are
distillable \cite{dist}. Class (b), on the other hand contains all the bound
entangled (BE) states \cite{bound,Pawel}. Indeed, two qubit entangled states
are distillable hence a BE state cannot contain two-qubit entanglement. A
known open question in this context is if class (b) is equivalent to the BE
states or if it is strictly larger. In Ref. \cite{bound} it was shown that a
state $\varrho $ is distillable iff for some number $k$ the state $\varrho
^{\otimes k}$ has two-qubit entanglement. Call such state k-copy pseudo
distillable (according to notation of Ref. \cite{DiVincenzo}). The question
if the set of BE states is equal to class (b) can be then rephrased as: does
k-copy pseudo distillability imply 1-copy pseudo distillability. In
principle it might happen that the property of having two-qubit entanglement
is not additive: 1 copy would not contain it, but two or more copies would.
For some Werner states there is strong evidence that this is the case \cite
{DiVincenzo,NPTLew}. In Ref. \cite{Gisin} a possible equivalence of the
considered sets was connected with some ``binarization'' of conditional
information in cryptography based on mixed quantum states.

In this context, our preconcurrence matrix allows for a simple argument
which shows that rank 2 states are either separable or 1-copy pseudo
distillable (for the original proof of non-existence of bound entangled
states of rank 2 see \cite{PawelSTT}).

\subsection{Rank-2 states are either separable or 1-copy pseudo distillable.}

Rank-2 states have $2\times 2$ preconcurrence matrices. A state which has a
decomposition where all the matrices are of the form 
\begin{equation}
C_{1}=\left[ 
\begin{array}{cc}
0 & x \\ 
x & 0
\end{array}
\right]  \label{C-1}
\end{equation}
is separable. A candidate for a non-separable and not 1-copy
pseudo-distillable state must have at least two essentially different
preconcurrence matrices. In a decomposition where one of the matrices is of
the form (\ref{C-1}), there must be another one 
\begin{equation}
\tilde{C}_{2}=e^{i\varphi }\left[ 
\begin{array}{cc}
a~e^{i\alpha } & b \\ 
b & -a~e^{-i\alpha }
\end{array}
\right]  \label{C-2}
\end{equation}
with all the parameters real and $a\neq 0$. This form is necessary since
otherwise it would be impossible for transformation (\ref{Con-sim}) to make
the diagonal of $\tilde{C}_{2}$ zero. Moreover, a simple phase adjustment in
the decomposition of the state can bring $\alpha $ and $\varphi $ to zero,
without changing $\tilde{C}_{1}$'s diagonal. With such an adjustment the
second matrix becomes 
\begin{equation}
C_{2}=\left[ 
\begin{array}{cc}
a~ & b \\ 
b & -a~
\end{array}
\right]  \label{C-2a}
\end{equation}
with both $a$ and $b$ real. Now, a change of the local bases which (up to a
normalizing factor) produces 
\[
C_{2}^{\prime }=C_{2}+i\left[ 
\begin{array}{cc}
0 & \left| x\right| \\ 
\left| x\right| & 0~
\end{array}
\right]
\]
shows that the state contains 2-qubit entanglement, i.e., it is distillable.
Indeed, $C_{2}^{\prime }$ is of form (\ref{C-2}) with real non-zero $a$ and
complex $b$. Such a matrix has two singular values of different moduli.
Consequently, no transformation (\ref{Con-sim}) can reduce its trace to
zero. This implies 2-qubit entanglement.

As a corollary to the above argument, one may notice that a rank-2 state is
separable iff there exists a 2-state decomposition of the state which
simultaneously diagonalizes all the $C_{\kappa }$ matrices so that all the
matrices are of the essentially the same form 
\begin{equation}
C_{\kappa }=\left[ 
\begin{array}{cc}
x_{\kappa } & 0 \\ 
0 & -x_{\kappa }
\end{array}
\right]  \label{Cq-diag}
\end{equation}

Indeed, if separability requires existence of a decomposition where,
irrespectively of the choice of the local bases, all the $C_{\kappa }$'s are
of form (\ref{C-1}), then transformation (\ref{Con-sim}) with 
\begin{equation}
U=U_{q}=\frac{1}{\sqrt{2}}\left[ 
\begin{array}{cc}
1 & 1 \\ 
-1 & 1
\end{array}
\right]  \label{U-q}
\end{equation}
transforms them into (\ref{Cq-diag}).

Analysis of separability of states of rank higher than two appears to be
more difficult. In particular, an attempt to follow Wootters' minimization
procedure for the expectation value of the concurrence's norm is not simple
since there is no guarantee that transformation (8) can diagonalize matrix 
$C $ (notice that the elements of $C$ are vectors while the elements of $U$
are numbers. One can, nevertheless, diagonalize $D=Tr_{\mathcal{H}_1}C
C^{\dagger}$. This leads to some simplifications in special cases, like when
diagonal $D$ implies diagonal $C$. Nevertheless, at the moment, we do not
have any general results for states of rank higher than 2.

\section{Biconcurrence.}

\label{sec-biconcurrence}

Bearing in mind the difficulties, one may try to look at the generalized
concurrence from a somewhat different perspective. For two qubits,
preconcurrence can be viewed as a bilinear form $C(\psi, \phi)$ which
distinguishes between product vectors and entangled vectors. It satisfies
the following crucial condition:

\textbf{Condition 1.} $C(\psi,\psi)=0$ if and only if $\psi$ is a product
vector.

In passing, one may note that a form which satisfies condition~1 cannot be
linear in one argument and anti-linear in the other, since a
linear-antilinear form can be written as 
\begin{equation}
C(\psi ,\phi )=\langle \psi |A\phi \rangle ,  \label{li-ali}
\end{equation}
where $A$ is a linear operator acting on space $\mathcal{H}$. However, form (%
\ref{li-ali}) which vanishes on all the product vectors, vanishes
everywhere, thus violating condition~1. Consequently, the form must be
bi-linear (or bi-antilinear, it does not matter which). In this context,
Wootters' concurrence defines a good form for $\mathcal{H}=C^{2}\otimes C^{2}
$. It reads 
\begin{equation}
C(\psi ,\phi )=\langle \tilde{\psi}|\phi \rangle 
\end{equation}
Wootters' preconcurrence matrix is then simply

\begin{equation}
C^{\mu\nu}(\varrho)=C(\psi_\mu,\psi_\nu)
\end{equation}

Unfortunately, as it follows from Ref. \cite{Rungta}, in higher dimensions
there does not exist a bi-linear form satisfying Condition 1. A possible way
to generalize Wootters concurrence can then be to look for a 4-argument form 
$B(\psi, \phi, \kappa, \theta)$ which would satisfy

\textbf{Condition $1'$.} $B(\psi )\equiv B(\psi ,\psi ,\psi ,\psi )=0$ iff $%
\psi $ is a product vector. A possible form satisfying Condition $1'$, linear
in two arguments and antilinear in the two other is closely related to
Rungta et al. concurrence \cite{Rungta} and to our preconcurrence matrix.
For instance, one can take a slightly simplified version of concurrence in 
\cite{Rungta} as a departure point, and define 
\begin{equation}
B(\psi )=-\langle \psi |I\otimes \Lambda (|\psi \rangle \langle \psi |)|\psi
\rangle .
\end{equation}
where $\Lambda $ is the positive map  used in the reduction criterion of
separability \cite{xor}: $\Lambda (A)=Tr(A)I-A$

One finds that $B(\psi )=1-\mathrm{Tr}\varrho ^{2}$, where $\varrho $ is a
reduction of $\psi $. It is then clear that $\mathbf{B}$ satisfies the
condition~$1'$. The corresponding bi-concurrence matrix is then 
\begin{equation}
B^{\mu \nu mn}=B(\psi _{\mu },\psi _{\nu },\psi _{m},\psi _{n})=-\langle
\psi _{\mu }|I\otimes \Lambda (|\psi _{\nu }\rangle \langle \psi _{m}|)|\psi
_{n}\rangle .
\end{equation}
After some algebra this can be rewritten as 
\begin{equation}
B^{\mu \nu mn}=\langle \psi _{\mu }|\psi _{\nu }\rangle \langle \psi
_{m}|\psi _{n}\rangle -Tr\biggl[[\psi _{\mu }]^{\dagger }[\psi _{\nu }][\psi
_{m}]^{\dagger }[\psi _{n}]\biggr]
\end{equation}
which is nothing else than a partial contraction of a product of
preconcurrence matrix with its complex conjugation. 
\begin{equation}
B^{\mu \nu ~mn}={\frac{1}{4}}\sum\limits_{i\wedge j,k\wedge l}C_{i\wedge
j;k\wedge l}^{n\nu }\cdot \left( C_{i\wedge j;k\wedge l}^{m\mu }\right)
^{\ast }
\end{equation}
Bi-concurrence is invariant under local unitary rotations of the state.
Changes in the state's decomposition, on the other hand, transform
bi-concurrence as follows 
\begin{equation}
\tilde{B}^{\mu \nu mn}=\sum_{\alpha ,\beta ,a,b}(U^{\mu \alpha })^{\ast
}(U^{ma})^{\ast }B^{\alpha \beta ab}U^{\nu \beta }U^{nb}.
\end{equation}
If we treat the matrix $\mathbf{B}$ as an operator acting on tensor product
of Hilbert spaces with Greek (Latin) indices for first (second) space, we
obtain 
\begin{equation}
\tilde{\mathbf{B}}=U^{\ast }\otimes U^{\ast }\mathbf{B}(U^{\ast })^{\dagger
}\otimes (U^{\ast })^{\dagger }  \label{eq-transf}
\end{equation}
One can see that the matrix $B$ contains the whole information about
possible separability of state $\varrho $. Moreover, irrespective of the
decomposition, the elements on the main diagonal of $B$ are real and
non-negative. Therefore, in terms of biconcurrence, separability is
equivalent to the existence of such a unitary $U$ that in eq. (\ref
{eq-transf})
\begin{equation}
tr(\tilde{\mathbf{B}})=0  \label{tr-cond}
\end{equation}
The lower-case $tr$ is here understood as the sum of the elements 
on the main diagonal: 
\begin{equation}
tr\tilde{\mathbf{B}}=\sum_{\mu }\tilde{B}^{\mu \mu \mu \mu }.
\end{equation}
Note that the elements $\tilde B^{\mu\mu\mu\mu}$ are always nonnegative.
Therefore it suffices to minimize (\ref{tr-cond})over unitaries $U$ 
and check whether the minimum vanishes. 

Within the picture of $B$ acting on  product  Hilbert space
one can express the condition as follows
\be
\min_UTr(U\otimes U P U^\dagger \otimes U^\dagger B)=0,
\ee
where $P=\sum_i|ii\rangle\langle ii|$ with $|ij\rangle$
being standard product basis. 

The condition (\ref{tr-cond}) seems to be quite simple, and we hope that 
it will lead to a more operational condition for separability.

\section{Conclusions.}

\label{sec-conclusion} In conclusion, we argue that the multidimensional
generalizations of concurrence which we have introduced in this contribution
put the question of determination of separability of bipartite quantum
states in a somewhat new perspective. 

First, we introduced a concept of preconcurrence matrix. The matrix was
designed to distinguish between the contributions to the entanglement which
embrace pairs of different two dimensional subspaces of the bipartite
system. In this way, our preconcurrence matrix contained all the information
necessary to identify separability of a given state. Nevertheless, its
dependence on the particular choice of the local basis made it rather
difficult to analyze in detail, but in a rather restricted class of cases. 

Therefore, we also generalized the concept of concurrence in another
direction and abandoned the requirement for it to be a second order object
in the state's ensemble. We arrived at the concept of biconcurrence matrix.
This matrix is of the fourth order in the state's ensemble, however, for
this price it is invariant under local unitaries. Biconcurrence can be
easily derived from a given bipartite state directly. It can also be
constructed by a suitable contraction out of our preconcurrence matrix. The
resulting separability condition is probably the easiest possible one from
the algebraic point of view.

Regarding a complete characterization of entanglement, on the other hand,
our generalizations of concurrence matrix may not be enough. The main reason
for this is that in order to specify the singular values of $\left[ \psi %
\right] $, in addition to the length of the preconcurrence defined in eq. (%
\ref{Con-pure norm}), one needs the lengths of all its tri-, ..., d-linear
analogues. We hope to return to this point in the nearest future.


P.B. is partially supported by Svenska Institutet, project ML2000. In
addition, P.B. acknowledges stimulating discussions with Richard Bonner and
Benjamin Baumslag. M.H., P.H. and R.H. are supported by Polish Committee for
Scientific Research, contract No. 2 P03B 103 16 and by UE project EQUIP,
contract No. IST-1999-11053.

\end{document}